\documentclass[prl,floats,aps,superscriptaddress,showpacs]{revtex4}
\usepackage{amsmath}
\usepackage{graphicx}
\newcommand{\be}{\begin{equation}}
\newcommand{\ee}{\end{equation}}
\newcommand{\ba}{\begin{eqnarray}}
\newcommand{\ea}{\end{eqnarray}}
\newcommand{\nsigma}{\mbox{\boldmath $\sigma$}}
\newcommand{\fex}{\frac{1}{\tau_s}\mbox{e}^{-T_s/T^{\ast}}}
\newcommand{\br}{{\bf r}}
\newcommand{\gam}{{\tau_b^{-1}}}
\newcommand{\bv}{{\bf v}}

\begin{document}
\title{{\bf Statistical mechanics of granular 
gases in compartmentalized systems}}

\author{U. Marini Bettolo Marconi}
\affiliation{Dipartimento di Fisica, Universit\`a di Camerino and
Istituto Nazionale di Fisica della Materia,
Via Madonna delle Carceri, 62032 , Camerino, Italy}
\author{A.Puglisi}
\affiliation{Dipartimento di Fisica, Universit\`a La Sapienza 
and INFM Center for Statistical Mechanics and Complexity,                 
Piazzale A. Moro 2, 00185 Roma, Italy}

%%%%%%%%%%%%%%%%%%%%%%%%%%%%%%%%%%%%%%%%%%%%%%%%%%%%%%%%%%%%%%%%%%%

\begin{abstract}
We study the behavior of an assembly of $N$ granular particles contained
in two compartments within a simple kinetic approach.
The particles belonging to each compartment 
collide inelastically with each other
and are driven by a stochastic heat bath. In addition, the
fastest particles can
change compartment at a rate which depends on their kinetic energy.
Via a Boltzmann velocity distribution 
approach we first study the dynamics of the model
in terms of a coupled set of equations for the populations
in the containers and their granular temperatures and find 
a crossover from a symmetric high-temperature phase to an asymmetric 
low temperature phase. Finally,
in order to include statistical fluctuations, we solve the model within the
Direct Simulation Monte Carlo approach.
Comparison with previous studies are presented. 
\end{abstract}
\maketitle

%%%%%%%%%%%%%%%%%%%%%%%%%%%%%%%%%%%%%%%%%%%%%%%%%%ti%%%%%%%%%%%%%%%%%
%%%%%%%%%%%%%%%%%%%%%%%%%%%%%%%%%%%%%%%%%%%%%%%%%%%%%%%%%%%%%%%%%%%
%\tightenlines
\section{Introduction}

In recent years considerable progress has been achieved in
the understanding of granular gases, i.e.
assemblies of moving inelastic solid grains mutually colliding
and loosing a little energy in each collision \cite{gases,jaeger}.
They exhibit a fascinating and 
rich phenomenology which comprehends clustering
\cite{goldhirsch,gold2,mcnamara}
spontaneous vortex formation \cite{baldassa}, and breakdown of kinetic
energy equipartition in granular mixtures \cite{Dufty,Ioepuglio}.
More fundamentally granular gases represent one of the prototypical
non equilibrium systems.

Major progress have been achieved by considering the simplest
situations, i.e. spatially homogeneous and time-independent systems.
In spite of that, inhomogeneities do not always represent a nuisance
or an undesired complication, but, on the contrary, they can be a
source of new insight. Within such a perspective, external fields have
been introduced on purpose in order to probe the inhomogeneous
behavior of granular gases.  The seminal work of Schlichting and
Nordmeier \cite{experiment} has stimulated a vivid interest in the
behavior of the so called compartmentalized systems. A box, whose base
moves periodically up and down, is separated by a vertical barrier
into two compartments which communicate through an orifice.  For
strong shaking the two halves are equally populated, as it would occur
in the case of a standard molecular fluid, whereas for weaker driving
the symmetry is spontaneously broken.  The resulting scenario
resembles that of an equilibrium thermodynamical phase transition,
with the population difference between the two compartments playing
the role of the order parameter and the driving intensity that of
temperature. Our understanding of the problem has increased
since then due to a series of studies. These range from new
experiments \cite{Lohse,Lohse1}, 
phenomenological flux models~\cite{Eggers,Droz}, hydrodynamic equations
and Molecular Dynamics simulation \cite{Brey}.

The scope of the present contribution is to show that by choosing a
simplified yet significant model of compartmentalized granular gas it
is possible not only to derive a set of equations describing the
evolution of the macro-state of the system, but also to obtain
information about its microscopic fluctuations.  The present approach
represents a bridge between the phenomenological level of
refs. \cite{Lohse,Lohse1,Droz} and the statistical mechanical level
\cite{dererumpallettarum}.

The present paper is organized as follows: in the first section we
define the model and introduce the statistical description of the
system, based on a Boltzmann equation for the distribution functions,
modified to take into account the stochastic driving.  At this stage
we follow the strategy of integrating out microscopic fluctuations
going from a microscopic description based on the Boltzmann equation
to a macroscopic level, where only the occupation numbers and the
granular temperatures of the two compartments, i.e. the first two
moments of the distribution function are taken into account. This is
equivalent to neglect inhomogeneities of the system at scales smaller
than the linear size of the compartments. To obtain qualitative
insight, we first try a Gaussian solution. In other words, we assume
that the velocity PDF's are Gaussians, whose normalizations and
variances (related to the occupation numbers of the compartments and
to their granular temperatures, respectively), can be determined by
means of a set of self-consistent differential equations. The analysis
shows the existence of a transition from a symmetric phase, where the
compartments are equally populated and are at the same granular
temperature, to an asymmetric phase, where the two compartments have
different properties.  Various predictions are made: we locate the
bifurcation point, obtain relations between the asymptotic values of
the temperatures and densities in the compartments, estimate the
characteristic times to observe the symmetry breaking, etc.  In the
second section, we relax the Gaussian hypothesis about the nature of
the velocity fluctuations and let the populations in the two
compartments to fluctuate as well. This is done by solving numerically
the full Boltzmann equation by means of the Direct Simulation Monte
Carlo (DSMC)
\cite{puglisi}. We notice deviations with respect to the
treatment of section I,  in particular near the critical
point.
Finally in the last section we present our conclusions. 

%%%%%%%%%%%%%%%%%%%%%%%%%%%%%%%%%%%%%%%%%%%%%%%%%%%%%%%%%%%%%%%%%%%%%%%%%5
%%%%%%%%%%%%%%%%%%%%%%%%%%%%%%%%%%%%%%%%%%%%%%%%%%%%%%%%%%%%%%%%%%%%%%%%%5

\section{The model}
Let us consider a system, composed of two compartments A and B of the
same volume, $V_A=V/2$, and containing $N_A$ and $N_B$ particles,
respectively. Particle pairs belonging to the same compartment may
collide inelastically loosing a fraction of their kinetic energy, but
conserving their total momentum.  The post-collisional velocities
($\bv_1^{\ast},\bv_2^{\ast}$) are determined by the transformation:

\begin{eqnarray}
{\bf v}^{\ast}_1={\bf
v}_1-\textstyle{\frac{1}{2}}(1+\alpha)({\bf
v}_{12}\cdot\hat{\nsigma})\hat{\nsigma}
\label{eq:uno}
\end{eqnarray}

where ${\bf v}_{12}={\bf v}_{1} -{\bf v}_{2}$ and $\alpha$ 
is the restitution coefficient.

In addition, we assume that 
all the grains are subject to the action of an external 
stochastic driving force and their motion between two successive
collisions is described by the following Ornstein-Uhlenbeck process:
\begin{equation}
\frac{{\rm d}{\bf v}_i}{{\rm d} t}= -\frac{1}{\tau_b} {\bf v}_i+ \hat{\bf{\xi}_i},
\label{eq:due}
\end{equation}

where $-\gam {\bf v}_i$ is a friction term and ${\bf {\xi}_i}$ a
Gaussian random acceleration, of zero average and variance given by:
\begin{equation}
\langle {\xi}_{i\alpha}(t) {\xi}_{j\beta}(t^\prime)\rangle =
2 \frac{T_b  m}{\tau_b} \delta_{ij}
\delta_{\alpha\beta}
\delta(t-t^\prime)
,
\end{equation}
where $T_b$ is a measure of the intensity of the driving.  Notice that
in the elastic case ($\alpha=1$), the average kinetic energy per
particle moving in a d-dimensional space is $K=d T_b/2$, which is
just the ideal gas value.  The simultaneous presence of frictional
dissipation and random ``kicks'' renders the kinetic energy of the
system stationary even in the absence of collisional dissipation.

To complete the model, we allow the particles contained in A (B) to
move into B (A) with a probability per unit time,
$\tau_s^{-1}$, provided their kinetic energy exceeds a given threshold, let
say $T_s=\frac{1}{2}m u_s^{2}$.
Such a mechanism schematizes the jump process by which
particles may pass from one compartment to the other by overcoming a
vertical barrier of height $h$ only if $\frac{1}{2}m v^{2}>mgh$.

We shall make the key assumption, dictated by mathematical
convenience, of ignoring the spatial gradients which characterize
experimental situations. To be specific we stipulate that within the
spatial domain, that represents a compartment, the system is
homogeneous.  The effect of the dividing vertical wall is schematized
by a selection rule of probabilistic nature which allows only the more
energetic particles to cross the barrier.

To make analytic progress we shall transform the stochastic evolution
equations for the velocities of the particles into a system of
deterministic equations for the distribution
functions. In order to achieve this description, 
we shall treat the collisions at the level of the Boltzmann molecular
chaos approximation. This approximation, widely
employed even in the case of granular systems, 
is equivalent to neglect correlations among the colliding
particles. 
 
In order to study the statistical evolution
of the system we assume that the single particle phase-space
distribution function 
$f(\br,\bv,t)$ satisfies the following properties:
%  in  compartment A, 
%the following normalization of the distributions is assumed:
\be
N_{A(B)}(t)=\int_{V_{A(B)}}{\rm d} {\bf r} \int{\rm d}\bv f({\bf r},
{ \bf v},t)
\ee
where $N_{A(B)}(t)$ is the average number of particles 
in compartment A (B) at instant $t$.
The average kinetic energy per particle, the granular temperature
$T_{A(B)}$, is defined as:
\be
T_{A(B)}(t)=\frac{1}{N_{A(B)}(t) d}
\int_{V_{A(B)}}{\rm d}\br \int {\rm d}\bv m \bv^2
f(\br,\bv,t)
\ee

In the following we shall make the assumption that
to describe the 
essential properties of the system, we do not need the
detailed information contained in $f(\br,\bv,t)$, 
but these can can be captured by the 
following coarse grained distributions obtained by
eliminating the ${\bf r}$ dependence of the original distributions
\begin{equation}
f_{A(B)}(\bv,t)=\frac{1}{V_{A(B)}}
\int_{V_{A(B)}}{\rm d} {\bf r} f({\bf r},
{ \bf v},t) 
\end{equation}
The change of $f_{A(B)}(\bv,t)$ is given by:
\be
\partial_t f_A(\bv_1,t)=I(f_A,f_A)+\frac{T_b}{\tau_b} 
\left(\frac{\partial}{\partial{\bf v}_1}\right)^2 f_A(\bv_1,t)+
\frac{1}{\tau_b} \frac{\partial}{\partial{\bf v}_1} f_A(\bv_1,t)-
\frac{1}{\tau_s}\theta(|\bv_1|-u_s)[f_A(\bv_1,t)-f_B(\bv_1,t)]
\label{eq:boltzmann}
\ee
where $\theta(x)$ is the Heaviside function.
The first term represents the change of the distribution
due to collisions, the second and the third terms are due
to the interaction with the heat-bath and the last describes the
population change due to the particles migrating from one 
compartment to the other~\cite{puglisi}.

In order to obtain an explicit expression
for the collision integral $I(f,f)$ in the case of 
inelastic hard spheres, we shall follow closely the derivation of
Ernst and Van Noije \cite{vannoije} and set:

\be
I(f_A,f_A) =\sigma^{d-1} \int {\rm d}{\bf v}_2 \int^\prime
{\rm d}\hat{\nsigma} ({\bf v}_{12}\cdot\hat{\nsigma})
\left\{\frac{1}{\alpha^2}f_A({\bf v}_1^{\ast\ast},t)f_A(
{\bf v}_2^{\ast\ast},t)-f_A({\bf v}_1,t)f_A({\bf v}_2,t)
\right\}
\ee
The prime on the $\hat{\nsigma}$ integration enforces 
the condition ${\bf v}_{12}\cdot\hat{\nsigma}>0$, 
where $\hat{\nsigma}$ is a unit vector
along the line of centers of the colliding spheres at contact, whereas
$\bv_i^{\ast\ast}$ represent the
precollisional velocities, which can be computed by inversion of
eq.~(\ref{eq:due}).

By integrating over the spatial coordinate and over the velocity
it is straightforward to obtain the equation for the rate of change
of the occupation numbers $N_A$ and $N_B$. 
It reads:
\be
\frac{{\rm d} N_A(t)}{{\rm d}t}=-\frac{V_A}{\tau_s}
\int{\rm d}\bv_1 [f_A({\bf v}_1,t)-
f_B({\bf v}_1,t)]\theta(|{\bf v}_1|-u_s).
\label{eq:salto}
\ee 

We see that when the typical time $\tau_s$ diverges, there is no
particle exchange between the two compartments, thus $N_A,N_B$ are
constant, and the temperatures reach a constant value, which depends
on the heat bath properties, the physical properties of the particles
and on their number.

In order to obtain the energy equation, instead,
we multiply both sides by $m\bv^2_1$ and integrate over 
coordinate and velocity space:
\be
\frac{{\rm d} [ N_A(t) T_A(t)]}{{\rm d}t}=
\frac{2}{\tau_b}N_A [T_b-T_A(t)]+
\frac{2}{d} \frac{N_A}{V_A}\sigma^{d-1}v_A \mu_2 N_A(t) T_A(t)
\ee
\be
-\frac{V_A}{\tau_s d}
\int{\rm d}\bv_1 m \bv_1^2 [f_A({\bf v}_1,t)-
f_B({\bf v}_1,t)]\theta(|{\bf v}_1|-u_s)
\label{eq:salto2}
\ee
where we have introduced, following ref. \cite{vannoije},
the non dimensional quantity $\mu_2$ :

\be
\mu_2=
-\frac{1}{v_A^3}\frac{V^2}{N^2}
\int{\rm d}\bv_1 \bv_1^2 I(f_A,f_A)=
\ee

$$
-\frac{1}{v_A^3}\frac{V^2}{N^2}
\int{\rm d}\bv_1 \bv_1^2
 \int {\rm d}{\bf v}_2 \int^\prime
{\rm d}\hat{\nsigma} ({\bf v}_{12}\cdot\hat{\nsigma})
\left\{\frac{1}{\alpha^2}f_A({\bf v}_1^{\ast\ast},t)f_A(
{\bf v}_2^{\ast\ast},t)-f_A({\bf v}_1,t)f_A({\bf v}_2,t)
\right\}
$$
with $v_{A(B)}^2=2 T_{A(B)}/m$. In the following we shall set $m=1$.
A more compact notation is achieved if one introduces
the collision frequency $\omega_A(t)$ and the
non dimensional spontaneous cooling rate $\gamma$:
\be
\omega_A=\frac{\Omega_d}{\sqrt{2 \pi}}\frac{N_A}{V_A}\sigma^{d-1}v_A \label{rate}
\ee
\be
\gamma=\frac{\sqrt{2 \pi}}{d \Omega_d}\mu_2
\ee
with ${\Omega_d}=\frac{2 \pi^{d/2}}{\Gamma(d/2)}$.  the surface area
of a d-dimensional unit sphere. In the remainder of the paper we shall
measure the temperature in units of $T_s$, time in units of $\tau_b$
and length in units of $\sigma$.
%We can rewrite:

%\be
%\frac{{\rm d} [ N_A(t) T_A(t)]}{{\rm d}t}=
%-2\gamma \omega_A N_A(t) T_A(t) 
%+\frac{2}{\tau_b} N_A [T_b-T_A]
%\ee
%\be
%-\frac{\tau_s^{-1}}{d}\int_{V}{\rm d}\br
%\int{\rm d}\bv_1 m \bv_1^2 [f_A({\bf v}_1,t)-
%f_B({\bf v}_1,t)]\theta(|{\bf v}_1|-u_s)
%\label{eq:salto3}
%\ee
To proceed further analytically we
choose $d=2$ and assume an approximate Gaussian form \cite{sonine}
of the distribution function:
\be
f_{A(B)}(\bv,t)=\frac{N_{A(B)}}{V_{A(B)}}\frac{1}
{\pi 2 T_{A(B)}}\exp(-\frac{\bv^2}{2 T_{A(B)}})
\ee

In this case one can evaluate explicitly the collision term
\cite{vannoije}:
\be
\mu_2=(1-\alpha^2)\frac{\pi^{d/2-1}}{\sqrt{2}\Gamma(d/2)}
\ee

\be
\gamma \omega_A=\sigma
(1-\alpha^2)\frac{N_A}{2 V_A}\sqrt{\frac{T_A}{m}}
\ee
We find:
\be
\frac{{\rm d} N_A(t)}{{\rm d}t}=\frac{1}{\tau_s}[N_B e^{-T_s/T_B}
-N_A e^{-T_s/T_A}]
\label{eq:exch0}
\ee 
where the temperature $T_s$ is given by $T_s=\frac{1}{2} m u_s^2$.
Notice that the right hand side of the equation above
represents the difference between the incoming flux and the outgoing
flux of compartment A. A similar expression has been chosen by
Droz and Lipowski \cite{Droz} on  phenomenological grounds. The model we study
allows us to obtain self-consistently the temperatures $T_A$ and $T_B$,
a feature which was not present in most of the studies
dedicated to compartmentalized systems\cite{extension}.  
In fact, we can write:

\be
\frac{{\rm d} [ N_A(t) T_A(t)]}{{\rm d}t}=
-\frac{2}{\tau_s}
[N_A (T_A+T_S)e^{-T_s/T_A} - N_B(T_B+T_S) e^{-T_s/T_B}]
-2 \gamma \omega_A N_A T_A
+\frac{2} {\tau_b}N_A( T_b-T_A) 
\label{eq:exch2}
\ee

\be
\frac{{\rm d} [ N_B(t) T_B(t)]}{{\rm d}t}=
-\frac{2}{\tau_s}
[N_B (T_B+T_S)e^{-T_s/T_B} - N_A(T_A+T_S) e^{-T_s/T_A}]
-2 \gamma \omega_B N_B T_B
+\frac{2} {\tau_b}N_B( T_b-T_B) 
\label{eq:exch3}
\ee

Notice that the Arrhenius type of behavior of the transition rate
$\tau_s e^{T_s/T_A}$ results from the combination of two
assumptions: a) the fact that only particles whose energy
exceeds the threshold $T_s$ can overcome the barrier; b)
the Gaussian ansatz for the velocity distribution functions.
The latter ingredient, although very convenient for numerical
work, could result too crude in some physically relevant
situations, because the dynamics represent a severe
probe of the extreme value
statistics of the system. It is well known that the velocity distribution
of driven and undriven granular assemblies might be characterized by fat 
velocity tails \cite{puglisi}.
%%%%%%%%%%%%%%%%%%%%%%%%%%%%%%% FIG 3 %%%%%%%%%%%%%%%%%%%%%%%%%%%%%%%%%
\begin{figure}[htb]
%\begin{center}
\centerline{\includegraphics[clip=true,width=8.cm, keepaspectratio,angle=0]
{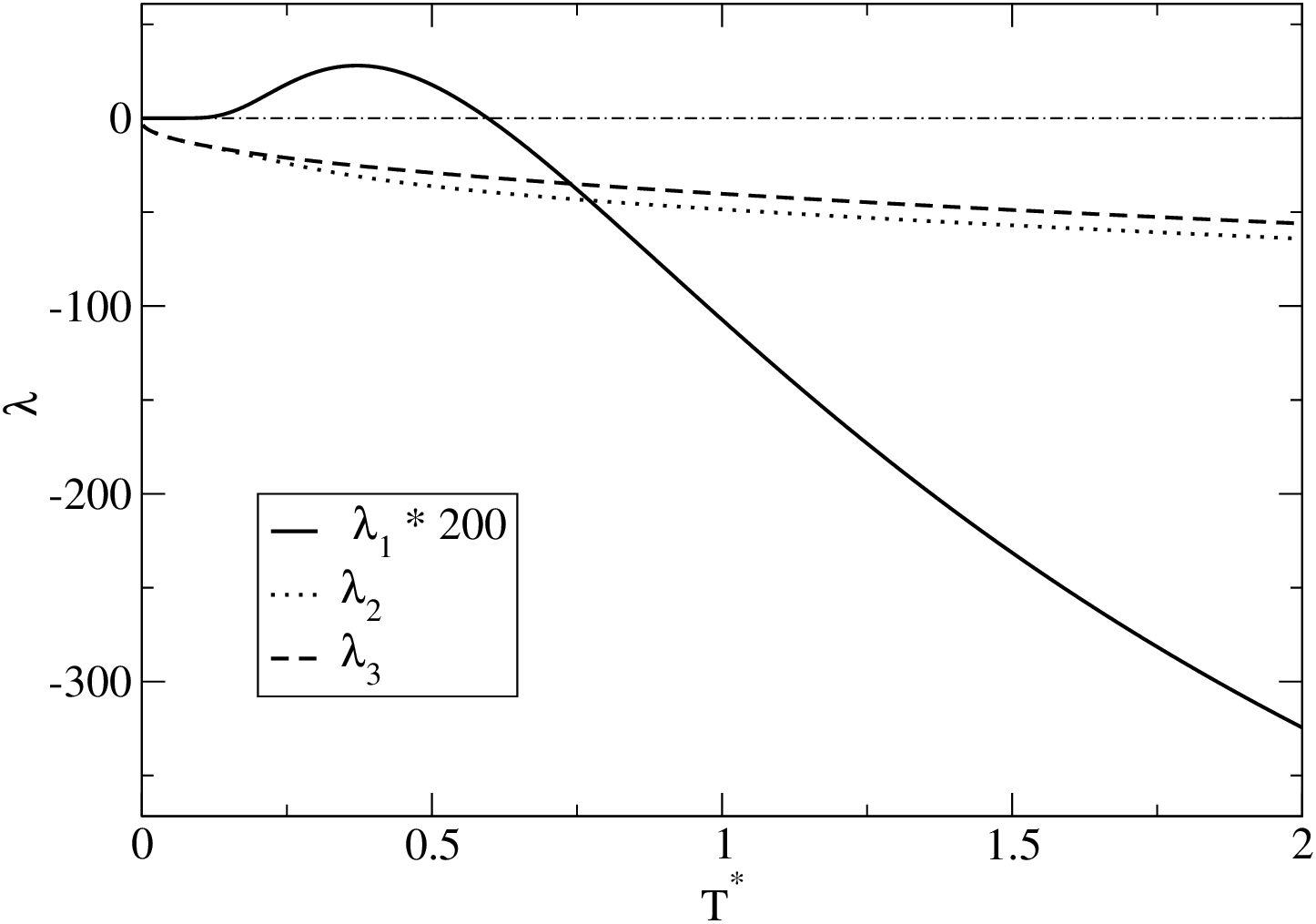}}
%\end{center}
\caption
{Variation of the eigenvalues of the dynamical matrix with 
respect to the granular temperature. Notice that 
below the critical temperature
the less negative 
eigenvalue displays a maximum. 
This is determined by the competition
between the inhibiting factor due to the wall 
and the collisional dissipation which favors a symmetry breaking.}
\label{3eigenvalues}
\end{figure}

%%%%%%%%%%%%%%%%%%%%%%%%%%%%%%% End FIG 3 %%%%%%%%%%%%%%%%%%%%%%%%%%%%%%%%%

\section{Mean field analysis}

It is convenient to rewrite the coupled equations as:

\begin{subequations} \label{eq:exch}
\begin{align}
\frac{{\rm d} N_A(t)}{{\rm d}t}&=\frac{1}{\tau_s}[N_B e^{-T_s/T_B}
-N_A e^{-T_s/T_A}]
\label{eq:exchA} \\
\begin{split}
N_A \frac{{\rm d} T_A(t)}{{\rm d}t}&=
-\frac{1}{\tau_s}
[2(N_A T_A e^{-T_s/T_A} - N_B T_B e^{-T_s/T_B})
+(N_A e^{-T_s/T_A} - N_B e^{-T_s/T_B})(2 T_s-T_A)] \\
&-2 \gamma \omega_A N_A T_A +\frac{2} {\tau_b}N_A( T_b-T_A) 
\label{eq:exchB} \end{split}\\
\begin{split}
N_B \frac{{\rm d} T_B(t)}{{\rm d}t}&=
-\frac{1}{\tau_s}
[2(N_B T_B e^{-T_s/T_B} - N_A T_A e^{-T_s/T_A})
+(N_B e^{-T_s/T_B} - N_A e^{-T_s/T_A})(2 T_s-T_B)] \\
&-2 \gamma \omega_B N_B T_B  +\frac{2} {\tau_b}N_B( T_b-T_B) 
\label{eq:exchC} \end{split}
\end{align}
\end{subequations}

Let us observe that in the temperature equations there
appear two types of fluxes: the first due to the unbalance of kinetic
energies, and the second due to 
and the population difference. We identify
the first with a heat conduction process and the second with a
particle diffusion process.

One sees by inspection that the choice
$N_A=N_B=N^{\ast}=N/2$ and $T_A=T_B=T^{\ast}$
represents a symmetric solution for all values of the
control parameters. The granular temperature of the
symmetric state, $T^{\ast}$, is
given by the nonlinear equation:
\be
T^{\ast} \left[1+
\tau_b\sigma
(1-\alpha^2)\frac{N^{\ast}}{2 V_A}\sqrt{\frac{T^{\ast}}{m}}\right]=T_b
\ee
In order
to ascertain the stability of such a symmetric solution we
assume $T_A=T^{\ast}+\delta T_A$,
$T_B=T^{\ast}+\delta T_B$ and $N_A=N^{\ast}+\delta N_A$
($\delta N_B=-\delta N_A$)
and expand the equations to linear order about
the fixed point $T^{\ast},N^{\ast}$ with the result:

\be
\delta\dot N_A=-\fex \left[2\delta N_A+\frac{ N^{\ast}T_s}
{(T^{\ast})^2}(\delta T_A-\delta T_B)\right]
\ee

%%%%%%%%%%%%%%%%%%%%%%%%%%%%%%%%%%%%%
\be
\delta\dot T_A=-\fex \left[2+\frac{T_s}{T^{\ast}}+\left(\frac{T_s}{T^{\ast}}\right)^2\right]
(\delta T_A-\delta T_B)-\left(3\gamma\omega^{\ast}+\frac{2}{\tau_b}\right)\delta T_A
-\frac{2}{N^{\ast}}\left[\fex (T^{\ast}+2 T_s)+\gamma\omega^{\ast}
T^{\ast}\right]\delta N_A
\ee

%%%%%%%%%%%%%%%%%%%%%%%%%%%%%%%%%%%%

\be
\delta\dot T_B=\fex \left[2+\frac{T_s}{T^{\ast}}+\left(\frac{T_s}{T^{\ast}}\right)^2\right]
(\delta T_A-\delta T_B)-\left(3\gamma\omega^{\ast}+\frac{2}{\tau_b}\right)\delta T_B
+\frac{2}{N^{\ast}}
\left[\fex (T^{\ast}+2 T_s)+\gamma\omega^{\ast}T^{\ast}\right]\delta N_A
\ee
%%%%%%%%%%%%%%% EIGENVALUES 

The eigenvalues of the associated $3\times 3$ matrix of coefficients
give the three  relaxation modes. By adding and subtracting the last
two equations the linear system factorizes into a decoupled equation
for the average value of the two temperatures ($(\delta T_A+\delta
T_B)/2$) with eigenvalue
$\lambda_3=-(3\gamma\omega^{\ast}+\frac{2}{\tau_b})$ and a system of
rank 2 involving the temperature difference and the occupation
number. The more negative of the remaining eigenvalues (we call it
$\lambda_2$) is of the same order as $\lambda_3$, while the second,
say $\lambda_1$, is smaller in absolute value and vanishes at the
``special'' temperature, $T_{cr}$.  This is obtained by solving the
transcendental equation:
\be
\tau_s\gamma\omega^{\ast}\left(\frac{T_s}{T_{cr}}-\frac{3}{2}\right)
-\frac{\tau_s}{\tau_b}
=2 e^{-T_s/T_{cr}}
\ee

Fig. \ref{3eigenvalues} shows that above the temperature
$T_{cr}$, $\lambda_1$ is negative, thus making the symmetric
solution stable, whereas for temperatures below it $\lambda_1$ is
positive. In such a case the symmetric fixed point is unstable and the
solution flows away, to an asymmetric fixed point. Hereafter, we shall
use the word ``critical temperature'' or ``critical line'' instead of
``special'' \cite{facciamoipedanti}.

Let us also notice that when the typical collision time
is much shorter than the characteristic times $\tau_c$ and $\tau_b$ the 
temperature $T_{cr} \to 2 T_s/3$, which represents an upper bound 
for such a quantity.

At this stage few comments are in order.  The transition results from
the competition between two effects: the diffusion due to the
``thermal'' agitation of the particles, i.e. the tendency to fill all
the available space, and the dissipation of energy during collisions
which favors clusterization.

%\subsection{Asymmetric solution}

When the external drive is sufficiently weak
there appears a second fixed point. This is found by imposing the
detailed balance:
\be
\frac{N_B}{N_A}=\frac{e^{-T_s/T_A}}{e^{-T_s/T_B}}
\label{balance}
\ee
and the energy balance
\be
-\frac{2}{\tau_s}
(N_A T_A e^{-T_s/T_A} - N_B T_B e^{-T_s/T_B}) 
-2 \gamma \omega_A N_A T_A
+\frac{2} {\tau_b}N_A( T_b-T_A)=0
\ee
\be
\frac{2}{\tau_s}
(N_A T_A e^{-T_s/T_A} - N_B T_B e^{-T_s/T_B})
-2 \gamma \omega_B N_B T_B
+\frac{2} {\tau_b}N_B( T_b-T_B)=0
\ee
A good approximation to the highest eigenvalue in such a case reads:
$$
\lambda_1=\frac{2}{3}  e^{-T_s/T^{\ast}} (\omega^{\ast}
\gamma(T_s/T_{\ast}-3/2)-1/\tau_b-2 e^{-T_s/T^{\ast}} ).
$$

%\subsection{Metastability of the solutions}

One observes two regions where the solutions are apparently metastable.
The first is the low temperature region, where a detailed inspection
shows that the symmetric state, in reality, is very weakly unstable.
From fig. \ref{3eigenvalues} we see
that at low temperatures ($T^{\ast}\to 0$) the positive 
eigenvalue $\lambda_1$ vanishes
exponentially as ${e^{-T_s/T^{\ast}}}$,
thus indicating that the symmetric solution might appear
stable if the observation time is finite.
In fact, the inter-well diffusion for $T_b\to 0$ is almost completely
suppressed by the Arrhenius factor.

The second region where the system displays genuine (meta)stability
occurs just above $T_{cr}$. A branch with $N_A \neq N_B$ and $T_A \neq
T_B$ is observed by integrating nmerically the system~\eqref{eq:exch}
by means of an Euler scheme. These asymmetric stationary states are
obtained by preparing the system in a subcritical configuration
($T^{ast}<T_{cr}$), and successively increasing the temperature above
$T_{cr}$. The width of the region is about a ten per cent of $T_{cr}$.
The observed hysteresis agrees qualitatively with the results relative
to the model of ref. \cite{Droz}.

%\subsection{Numerical solutions of the mesoscopic equations}
%%%%%%%%%%%%%%%%%%%%%%%%%%%%%%%%%%%%%%%%%%%%%%%%%%%%%%%
%Again we find a high temperature regime 
%separated by a low temperature regime by a critical
%point. The phenomenology is quite similar to that
%of previous work in the literature, although in the present
%approach the mesoscopic transition rates are derived from the
%underlying microscopic model via the Boltzmann
%equation.

\begin{figure}[htb]
\centerline{\includegraphics[clip=true,width=6.cm, keepaspectratio,angle=0]
{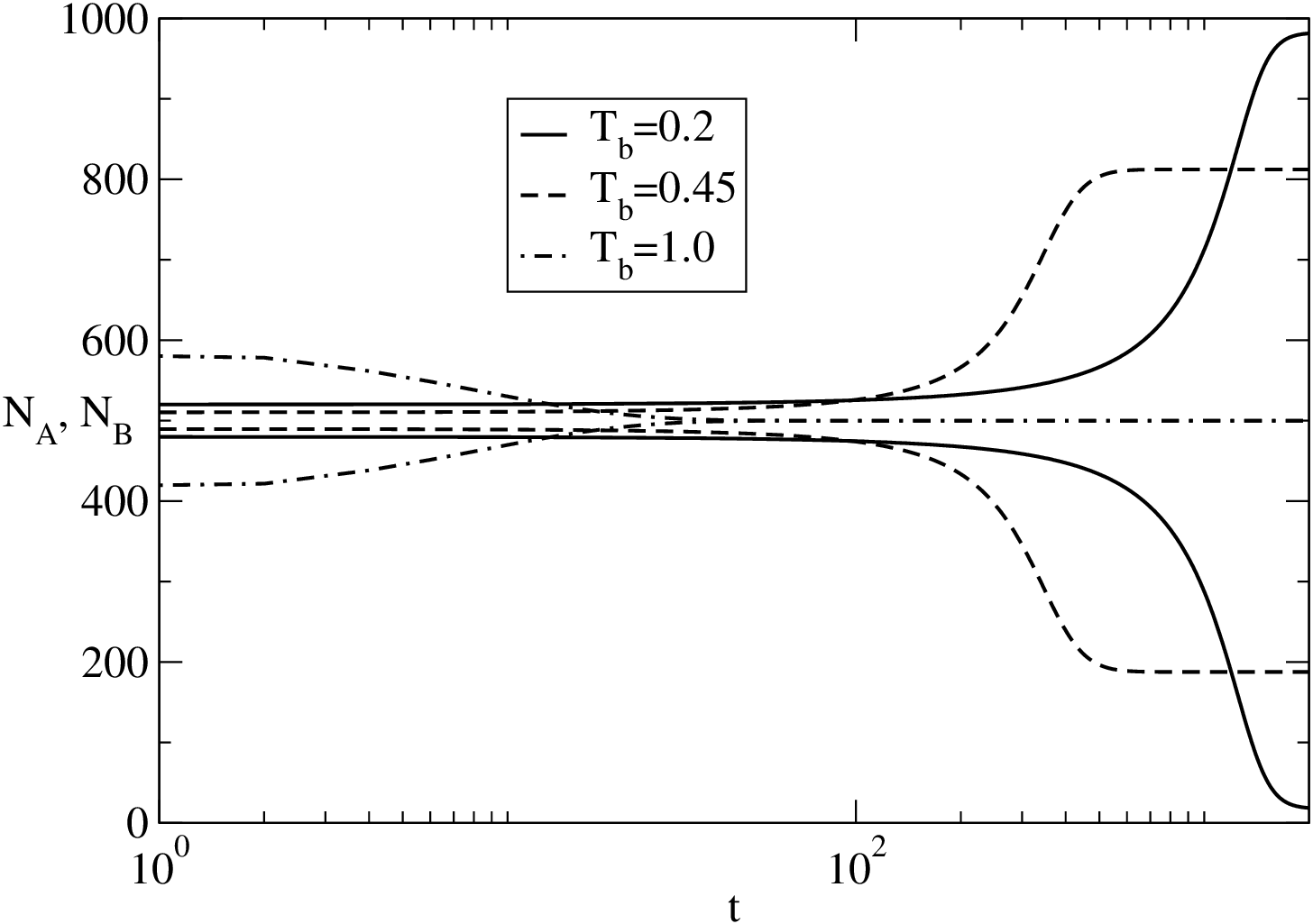}}
\caption
{Evolution of the 
populations in the two compartments versus time,
for three different choices 
of the heat bath temperature, $T_b=0.2, 0.3, 1$.
The remaining
parameters are: $N=1000$, $T_s=1$, $V=100$, $\sigma=1$, $\tau_b=1$,
$\tau_s=0.5$, $\alpha=0.7$.}
\label{occupations2}
\end{figure}

\begin{figure}[htb]
\centerline{\includegraphics[clip=true,width=6.cm, keepaspectratio,angle=0]
{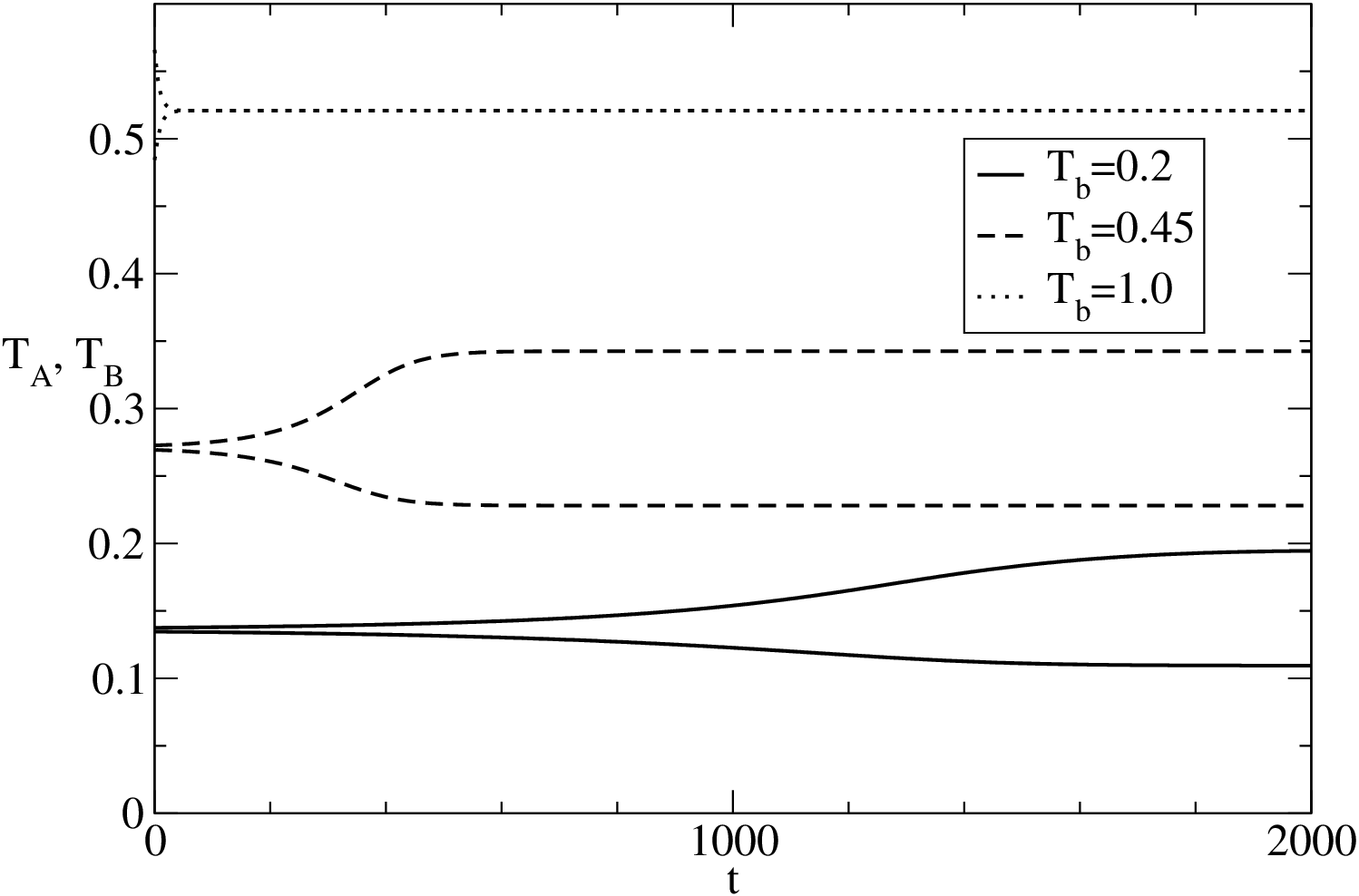}}
\caption
{ Granular temperatures in the two compartments
corresponding to the evolution of \ref{occupations2}.}
\label{anatempera}
\end{figure}

The numerical solutions of the coupled equations is displayed in figs.
\ref{occupations2} and \ref{anatempera}.
We observe that a perturbation about the symmetric solution is re-adsorbed
for $T^{ast}>T_{cr}$, whereas for temperatures below  $T_{cr}$ the 
perturbation grows initially at an exponential rate, before
saturating about a finite value. Let us notice that 
at low temperatures, due to the presence of the Arrhenius factor, 
the saturation
process occurs extremely slowly, a phenomenon which is
not related to the slowing down which occurs only in
the vicinity of $T_{cr}$.
%%%%%%%%%%%%%%%%%%%%%%%%%%%%%%% FIG occupation %%%%%%%%%%%%%%%%%%%%%%%%%%%%%%%%
%%%%%%%%%%%%%%%%%%%%%%%%%%%%%%%%%%%%%%%%%%%%%%%%%%%%%%%

%%%%%%%%%%%%%%%%%%%%%%%%%%%%%%% FIG  temperatures
%%%%%%%%%%%%%%%%%%%%%%%%%%%%%%%%%

%%%%%%%%%%%%%%%%%%%%%%%%%%%%%%%%%%%%%%%%%%%%%%%%%%%%%%%%%%%%%%

\begin{figure}[htb]
\centerline{\includegraphics[clip=true,width=6.cm, keepaspectratio,angle=0]
{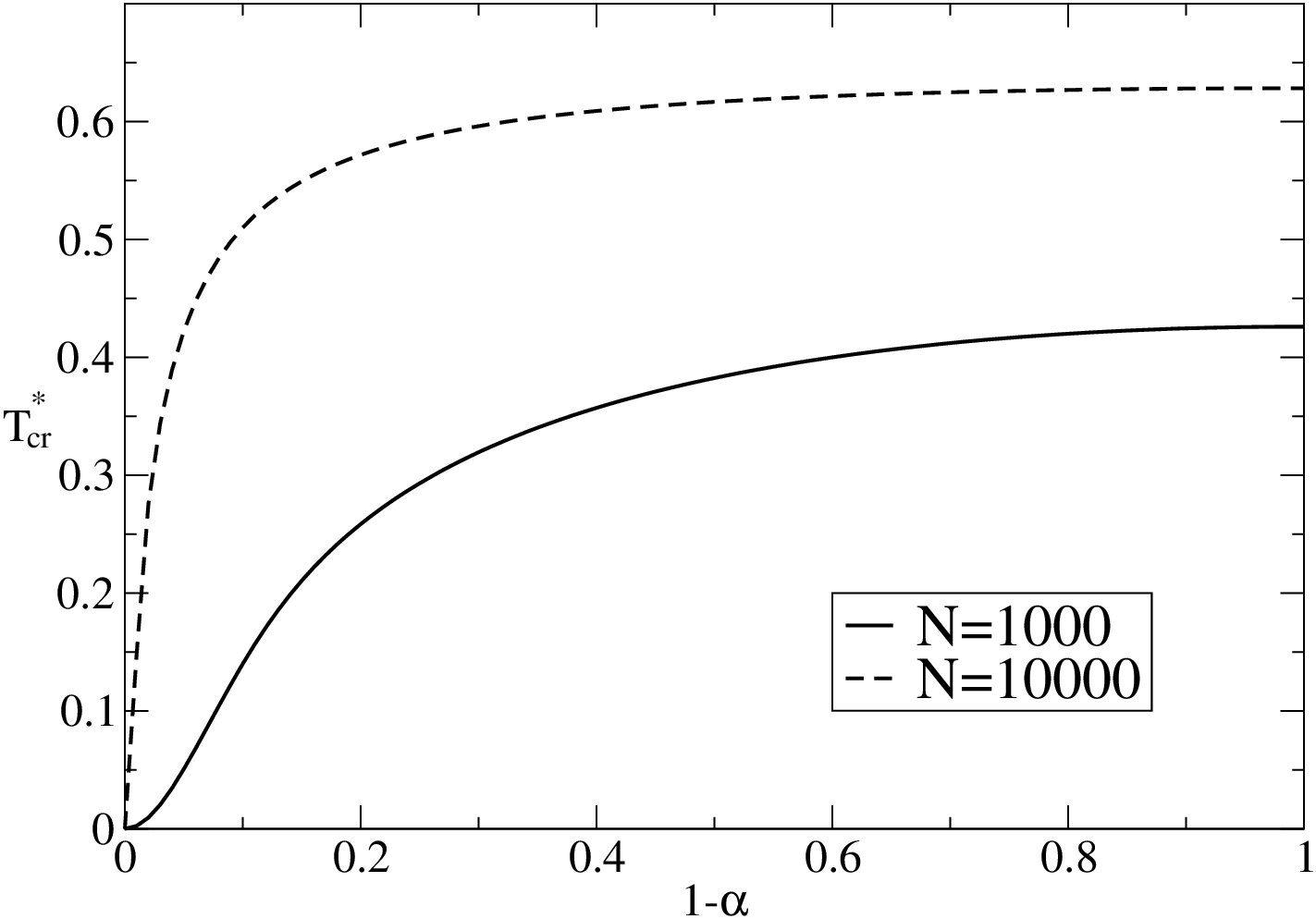}}
\caption
{Critical line for two different choices of the total number of particles
1000 (below) and 10000 (above) according
to the mean field theory. 
Vertical axis temperature, horizontal
$(1-\alpha)$. 
The line represents the granular critical temperature as the inelasticity
$(1-\alpha)$ varies from $0$ to $1$. Below the critical line
the symmetric solution is unstable, above it is stable.
The remaining parameters are the same as in fig.\ref{occupations2}.}
\label{lineacritica}
\end{figure}

In fig. \ref{lineacritica} we display the line of critical points 
$T_{cr}$ as a function of the inelasticity $(1-\alpha)$ for two different
values of the total number of particles. 
Above the line the solutions are symmetric ($N_A=N_B$), below
are asymmetric.
Let us observe that for
$\alpha=1$ one finds $T_{cr}=0$, because there is no clustering 
instability in 
systems of elastic particles. The smaller the value of $\alpha$
the higher the value of the critical temperature. We also notice
that the slope of the critical line is non negative. When the 
total number of particles decreases, also the critical temperature
decreases, because of the lower collision rate.

\begin{figure}[htb]
\centerline{\includegraphics[clip=true,width=6.cm, keepaspectratio,angle=0]
{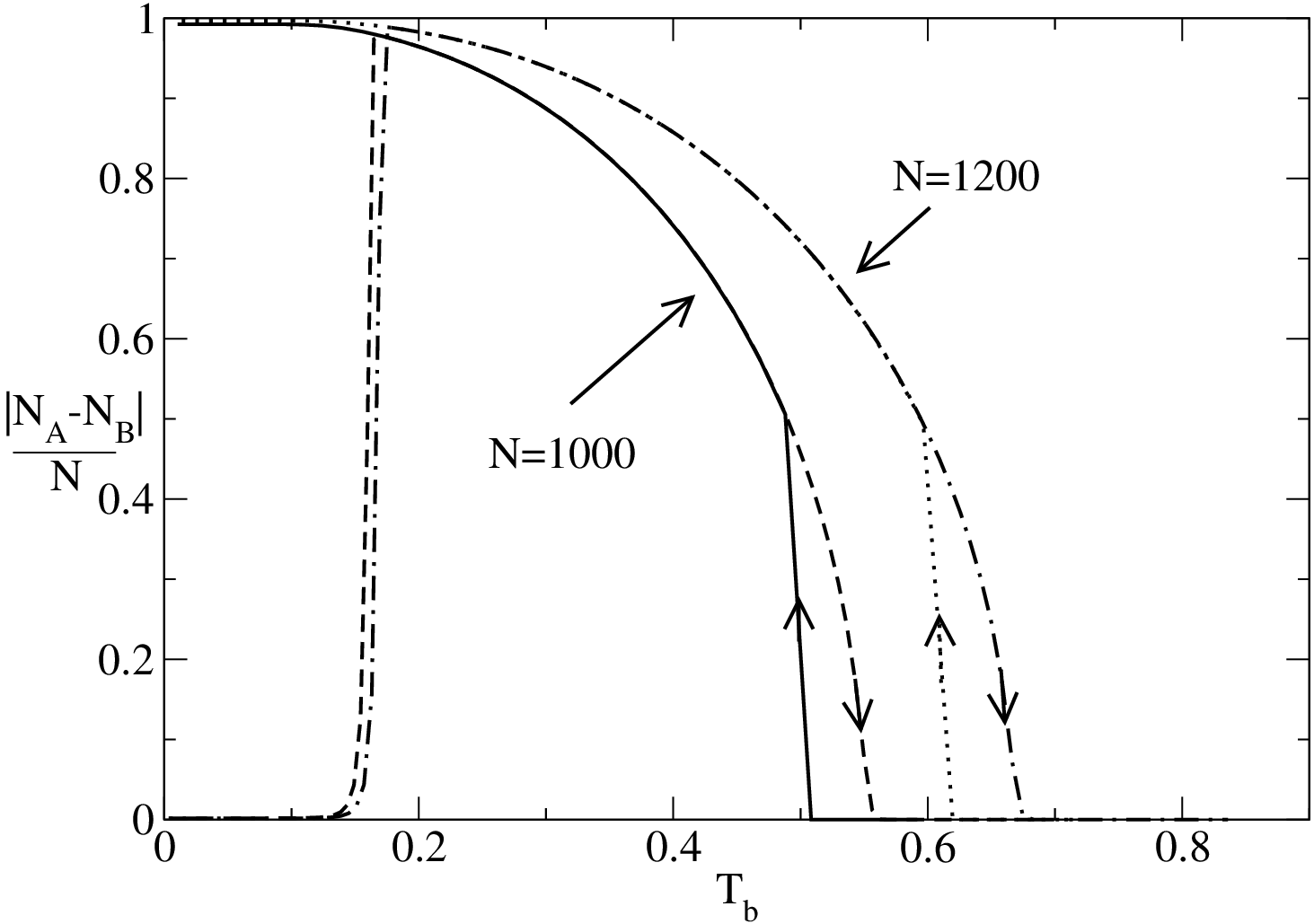}}
\caption
{Asymmetry parameter vs. $T_b$ plotted for two different choices of
the total number of particles: 1000 and 1200. The remaining parameters
are the same as in fig.\ref{occupations2}. Notice the difference
between the two curves and the presence of the hysteresis. }
\label{lineacoex}
\end{figure}

Figure \ref{lineacoex} illustrates the behavior of the order
parameter $\epsilon=|N_A-N_B|/N$ versus the heat bath temperature
for two different values of the total number of particles.

Let us notice that when particles are added to the system
the critical point shifts up-wards. This explains why the steady number
of particles in the less populated compartment decreases when more particles 
are added \cite{Brey}.

%%%%%%%%%%%%%%%%%%%% Fig Critical line %%%%%%%%%%%%%%%%%%%%%%%%%%%%%%%%%

%%%%%%%%%%%%%%%%%%%%%%%%%%%%%%% End FIg critical line%%%%%%%%%%%%%%%%%%%%%%%

%%%%%%%%%%%%%%%%%%%% Fig Coexistence line %%%%%%%%%%%%%%%%%%%%%%%%%%%%%%%%%
%%%%%%%%%%%%%%%%%%%%%%%%%%%%%%% End FIg coex line%%%%%%%%%%%%%%%%%%%%%%%

Finally,
we comment that, within our perspective, the approach of Lohse
et al. \cite{Lohse} and of Lipowski and Droz \cite{Droz}
is equivalent to an adiabatic approximation for the temperature
variables. In their cases, the two granular temperatures
are assumed to be slaved by
the occupation number variables. In other words,
one postulates that
the temperatures depend on the
instantaneous values of the occupation numbers.  
 We implemented this idea within our approach, but we found
a large discrepancy with the previous 
approximation. In reality, the temperature and the occupation
variable vary on the same time scale and no slaving principle 
seems to occur.

%%%%%%%%%%%%%%%%%%%%%%%%%%%%%%%%%%%%%%%%%%%%%%%%%%%%%%%%%%%%%
%%%%%%%%%%%%%%%%%%%%%%%%%%%%%%%%%%%%%%%%%%%%%%%%%%%%%%%%%%%%%%%%%%%%%%%%
%%%%%%%%%%%%%%%%%%%%%%%%%%%%%%%%%%%%%%%%%%%%%%%55555555

\section{Monte Carlo simulations}

In deriving the evolution equations of the 
previous section we have implicitly stipulated that the 
values of the occupation 
number and of the kinetic energy per particle in each box have
narrow distributions
around their mean values and that
these quantities are sufficient to characterize the state of
the system. Do statistical fluctuations 
play any role or even modify the picture presented above?
In the present section, which is not intended  
to be exhaustive, we solve the model of section I 
by means of the DSMC method.
The motivation is twofold: we want to  
validate the picture and the approximations introduced above
and unveil some phenomena which are not accounted for
by a mean field description.

We simulated two ensembles of $N_A$
and $N_B$
particles, respectively, subject to a Gaussian forcing, viscous friction and
inelastic collisions. In addition,
the particles of high energy can change compartment with
probability per unit time $\tau_s^{-1}$.

The scheme consists of the following ingredients:

\begin{itemize}

\item
time is discretized, i.e. $t=n\cdot dt$ 

\item
update all the velocities to simulate the
random forcing and the viscous damping:

\begin{equation}
v_i^{\alpha}(t+dt)=
v_i^{\alpha}(t)e^{-\frac{dt}{\tau_b}}+\sqrt{T_b\left(1-e^{-\frac{2dt}{\tau_b}}\right)}W(t)
\label{discre}
\end{equation}
where $W(t)$ is a normally distributed deviate with zero mean and unit
variance.

\item
At every time step and for both compartments, a collision step
is performed, i.e. an adequate number of randomly chosen pairs of
velocities are updated with the collision rule~\eqref{eq:uno}: the
pairs are chosen with a probability proportional to their relative
velocity and normalized in order to have a mean collision frequency as
calculated in eq.~\eqref{rate}.

\item
Place in the other compartment with probability $dt/\tau_s$ 
particles with kinetic energy greater than
$T_s$.

\item 
Change the time counter $n$ and restart.

\end{itemize}
 
To summarize, at every step each particle experiences a gaussian kick
and receives energy from the bath, but dissipates a fraction of its
kinetic energy by collision and by damping.  With respect to previous
DSMC simulations of granular gases particles can migrate to another
compartment, whenever their energy exceeds a fixed threshold. No
packing effects are included.

The parameters chosen in the simulation are 
$T_s=1$, $V=100$, $\sigma=1$, $\tau_b=1$,
$\tau_s=0.5$, $\alpha=0.7$, whereas the total number of particles
$N$ and $T_b$ have been varied. 

%%%%%%%%%%%%%%%%%%%%%%%%%%%%%%%%%%%%%%%%%%%%%%%%%%%%%%%%%5
%%%%%%%%%%%%%%%%%%%%%%%%%%%%%%% FIG  %%%%%%%%%%%%%%%%%%%%%%%%%%%%%%%%%
\begin{figure}[htb]
\centerline {\includegraphics[clip=true,width=10cm,keepaspectratio]
{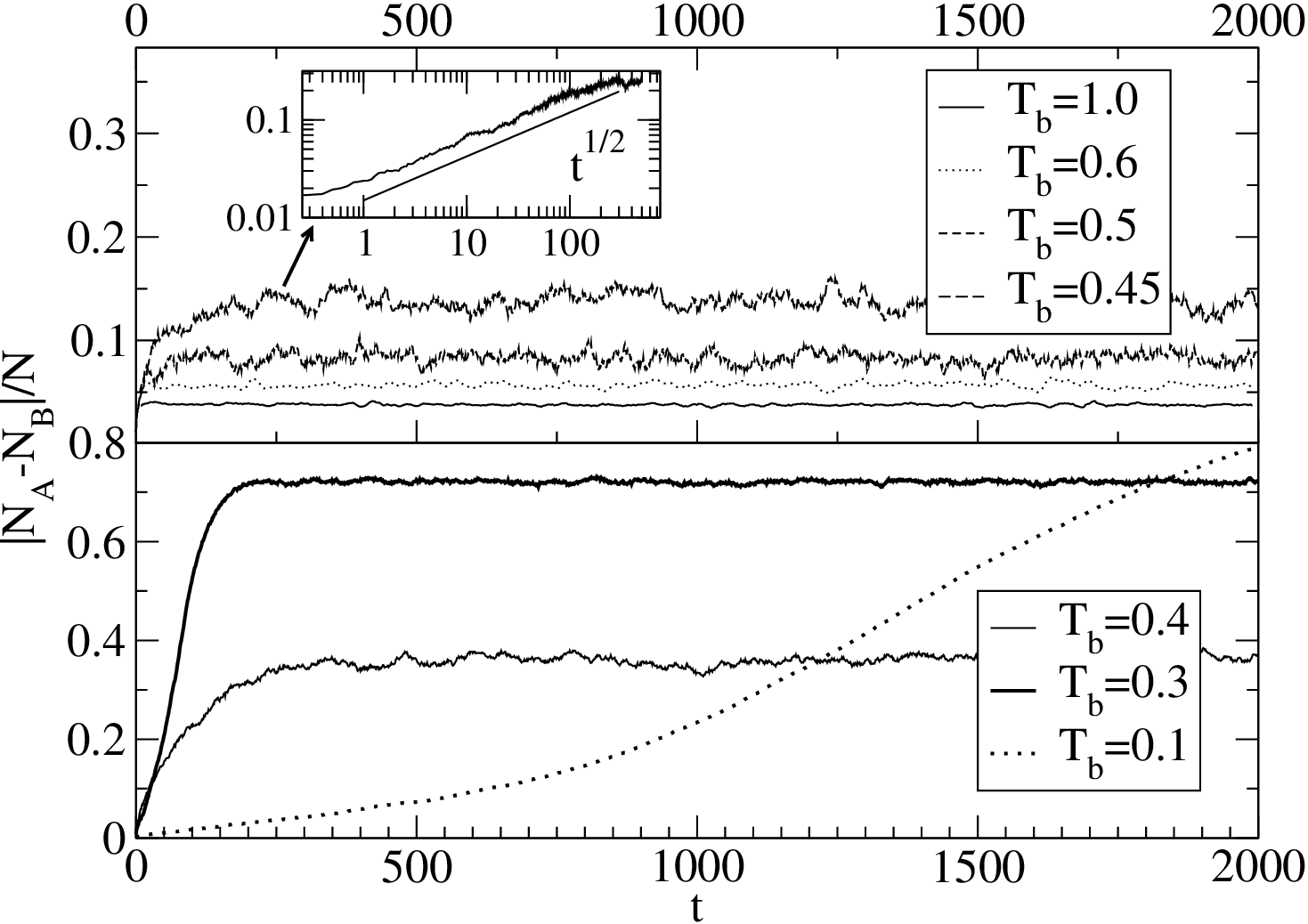}}
\caption
{Evolution of the order parameter
versus time. $N=1000$, but $T_b$ is varying, while the other
parameters are: $T_s=1$, $V=100$, $\sigma=1$, $\tau_b=1$,
$\tau_s=0.5$, $\alpha=0.7$.}
\label{fig_order}
\end{figure}
%%%%%%%%%%%%%%%%%%%%%%%%%%%%%%% End FIG %%%%%%%%%%%%%%%%%%%%%%%%%%%%%%%%%

%%%%%%%%%%%%%%%%%%%%%%%%%%%%%%% FIG %%%%%%%%%%%%%%%%%%%%%%%%%%%

\begin{figure}[htb]
\centerline{\includegraphics[clip=true,width=10.cm, keepaspectratio]
{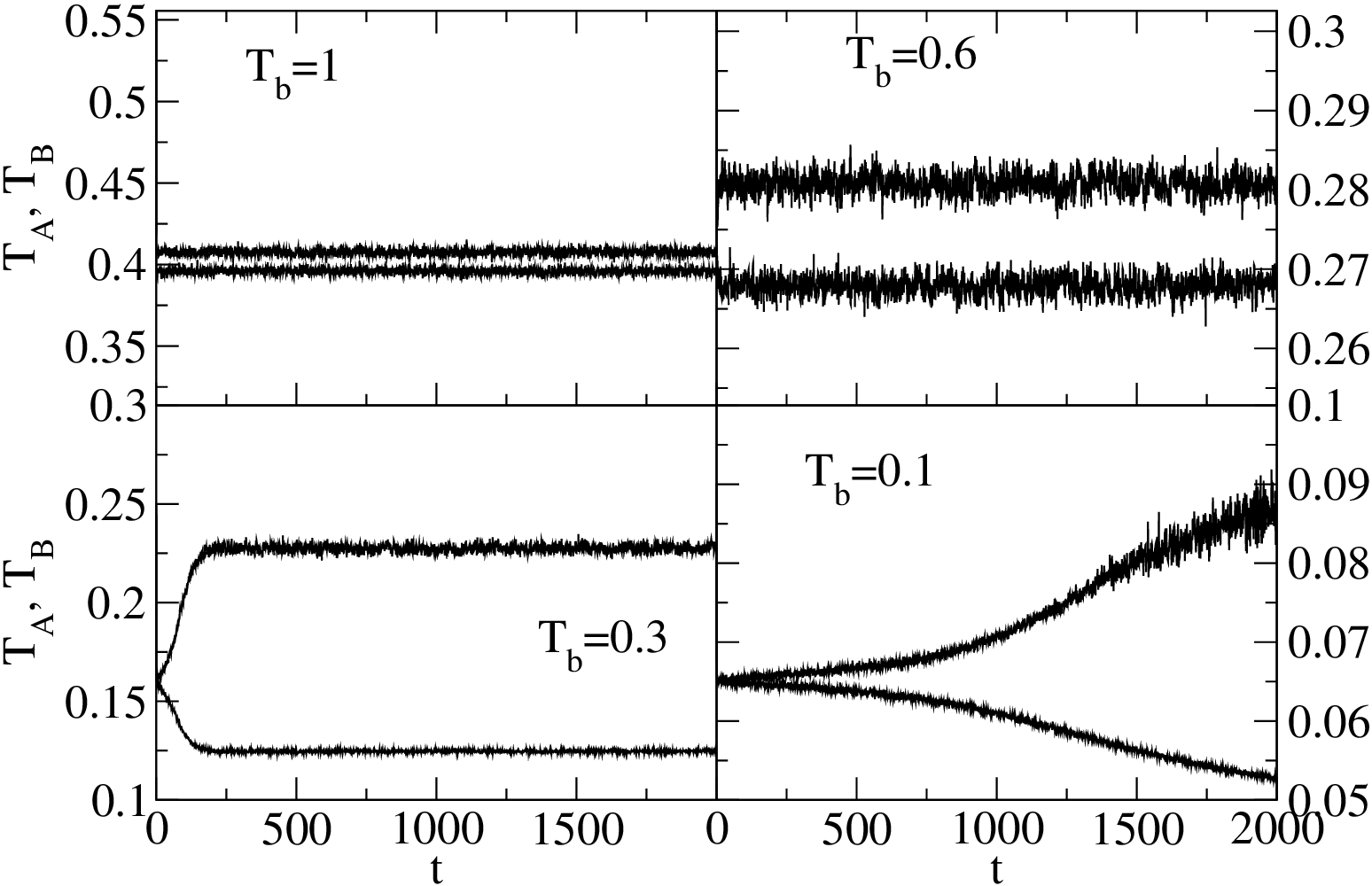}}
\caption
{Temperature evolution for various values of the heat bath temperature
The other parameters are the same as in fig.~\ref{fig_order}}
\label{fig_temperatures}
\end{figure}
%%%%%%%%%%%%%%%%%%%%%%%%%%%%%%% END FIG %%%%%%%%%%%%%%%%%%%%%%%%%%%

%\subsection{Order parameter and distributions}

First, we analyze the evolution of the order parameter
$|N_A-N_B|/N$, when the system is prepared in a symmetric configuration.
The order parameter displays a behavior similar to that obtained by 
means of the mean
field theory. In addition, it displays fluctuations around its asymptotic
value. 
In fig.~\ref{fig_order}, an average over $200$ realizations of the
evolution of the order parameter $|N_A-N_B|/N$ is shown. The numerical
simulations are in agreement with 
the results of the mean field theory. For $T_b
> 0.45$ the system remains substantially homogeneous. For $T_b < 0.45$
the homogeneous state becomes unstable and the stable configuration
that is reached is strongly asymmetric. 

Interestingly,
we observe a much slower growth (see inset) near the critical temperature,
$T_{cr} \sim 0.45$. The same kind of phenomenon appears at low
temperatures ($T<T_s$), because the transitions from one compartment
to the other represent very rare events.

In figure~\ref{fig_temperatures} the temperature evolution $T_A(t)$
and $T_B(t)$ (after averaging over $200$ realizations and relabeling the
compartments in such a way that $A$ is always the most populated) are
displayed. The plateau values of $T_A$ and $T_B$ verify (with small
deviations, not larger than $10\%$) the equation~\eqref{balance}.

%%%%%%%%%%%%%%%%%%%%%%%%%%%%%%% FIG %%%%%%%%%%%%%%%%
\begin{figure}[ht]
\centerline {\includegraphics[clip=true,width=8cm,keepaspectratio]
{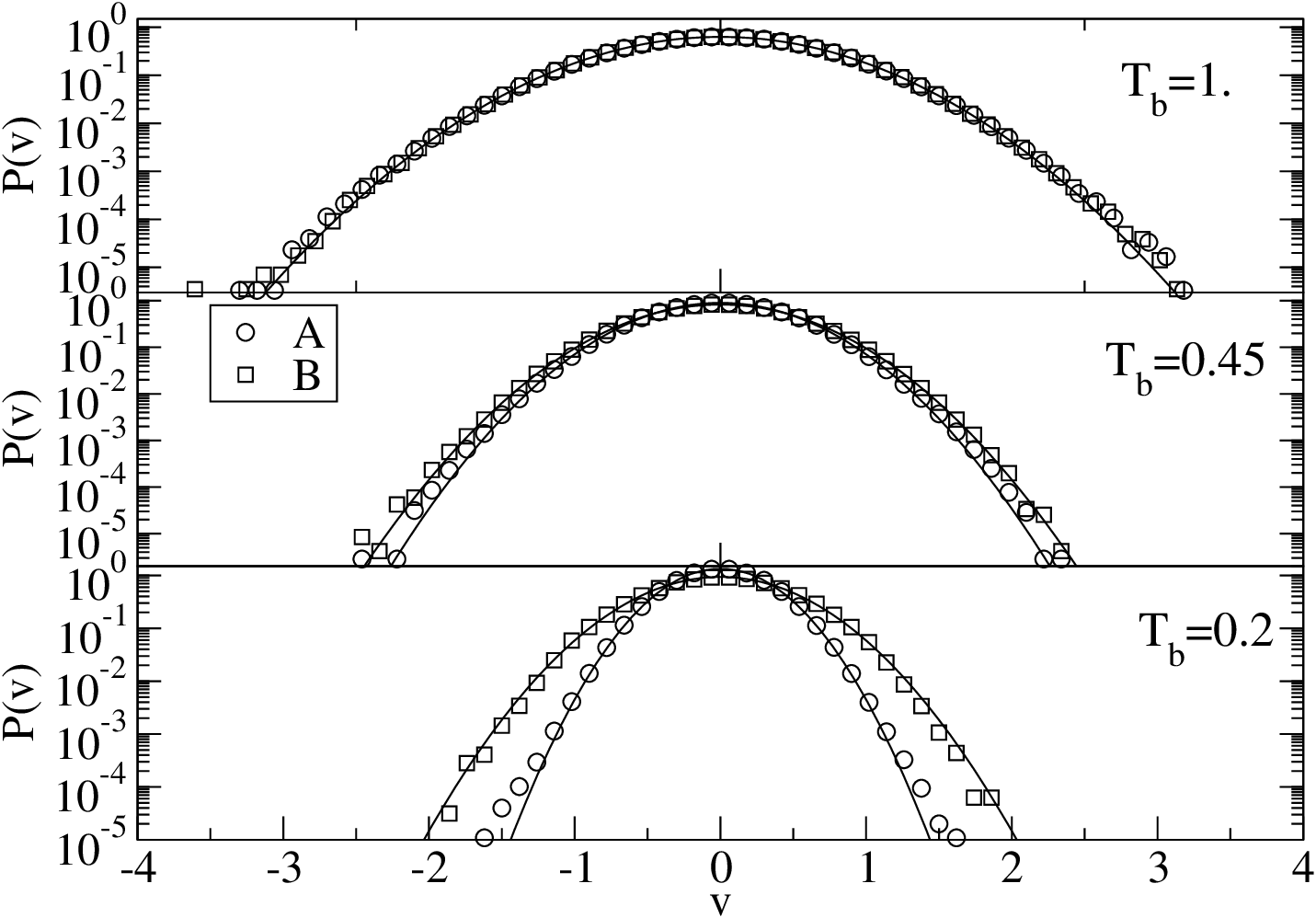}}
\caption{
Velocity distribution functions in each compartment for 
two different
values of $T_b$ and $N=1000$. 
The other parameters are the same as in fig.~\ref{fig_order}.
The lines represent Maxwellian velocity distributions, whose
variances have been determined from fits of the numerical data.} 
\label{fig_v}
\end{figure}
%%%%%%%%%%%%%%%% End crit.eps%%%%%%%%%%%%%%%%%%%%%

The velocity fluctuation can be appreciated by considering
the velocity PDF's of the two compartments
fig.~\ref{fig_v}. Appreciable deviations from a Gaussian have not
been detected
observed in all cases considered (neither above, neither below, nor in the
proximity of the critical temperature).

%%%%%%%%%%%%%%%%%%%%%%%%%%%%%%% FIG %%%%%%%%%%%%%%%%
\begin{figure}[ht]
\centerline {\includegraphics[clip=true,width=8cm,keepaspectratio]
{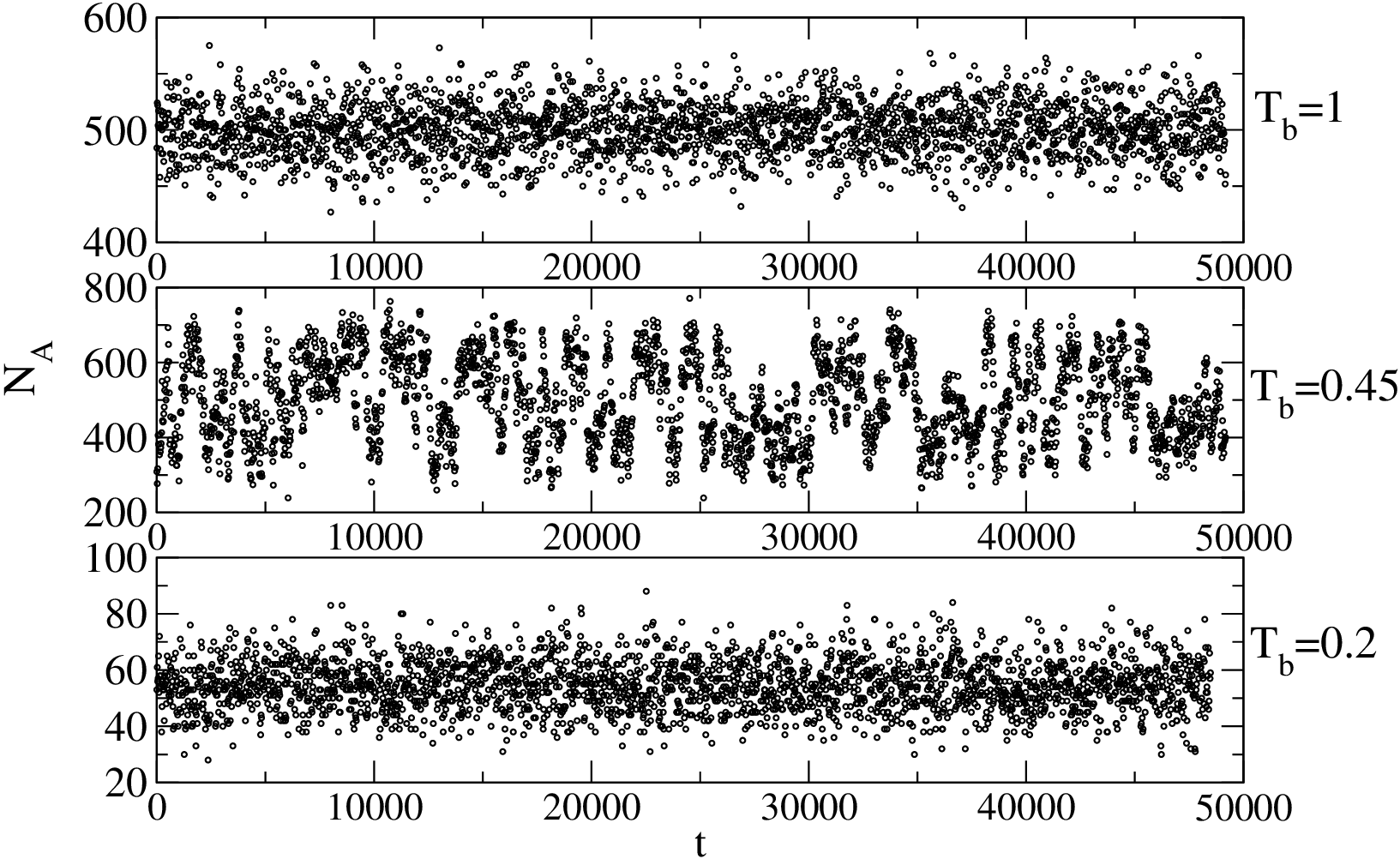}}
\caption{Fluctuations of the occupation number $N_A(t)$ vs time, when
the system is in its statistically stationary state for 
three different choices of $T_b$. 
The other parameters are the same as in fig.~\ref{fig_order}} 
\label{fig_n}
\end{figure}
%%%%%%%%%%%%%%%% End FI%%%%%%%%%%%%%%%%%%%%%%%%%%%%%%% FIG %%%%%%%%%%%%%%%%

\begin{figure}[ht]
\centerline {\includegraphics[clip=true,width=8cm,keepaspectratio]
{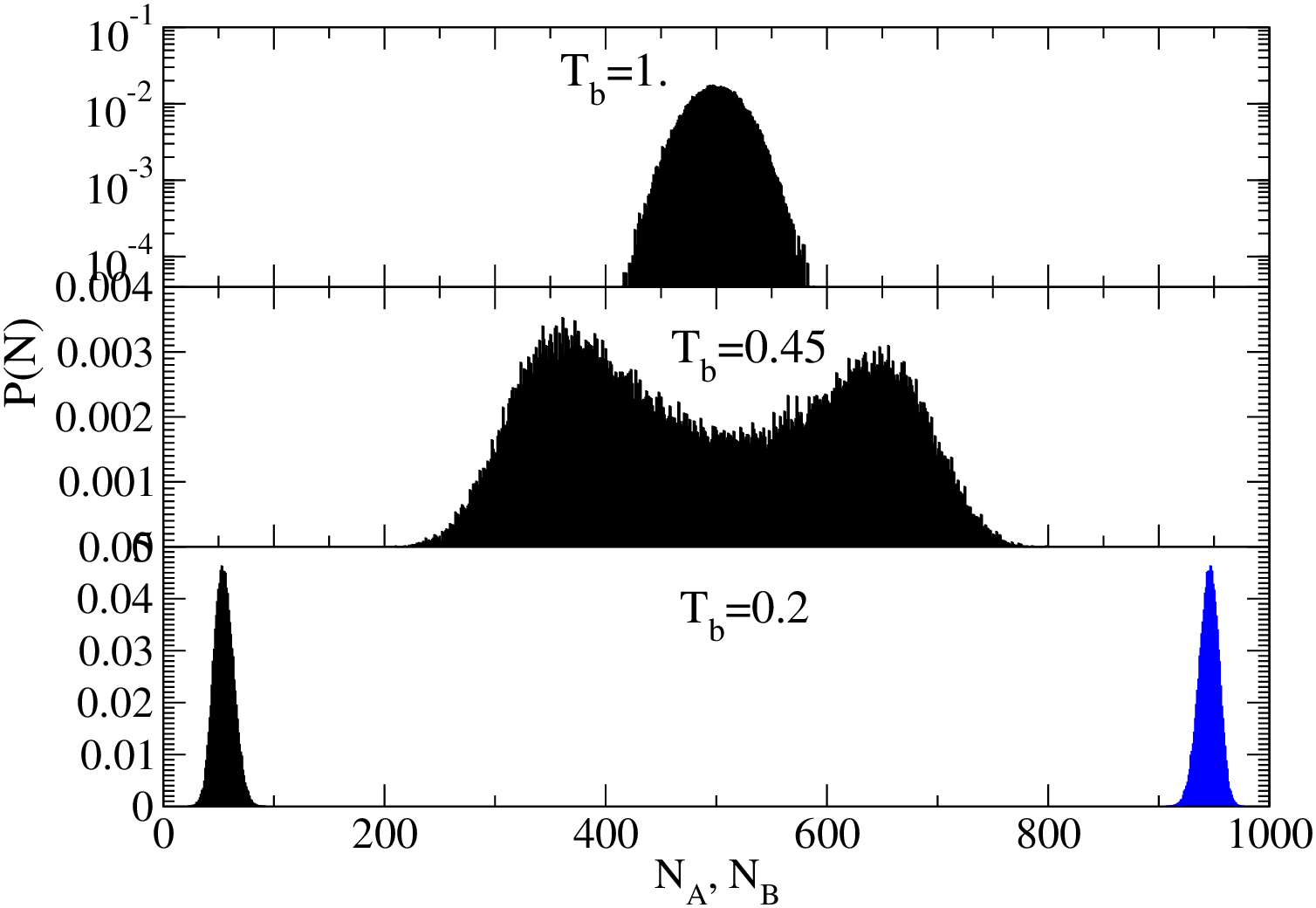}}
\caption{Probability distribution functions  of
the occupation numbers in a single well (top and middle panel,
respectively) and in the two wells (bottom panel), referred to a
system of $N=1000$ particles. 
The other parameters are the same as in fig.~\ref{fig_order}}
\label{fig_dn}
\end{figure}

The order parameter, instead, shows less trivial fluctuations. A typical
trajectory is shown in fig.~\ref{fig_n}.
 The associated distribution is shown
in fig.~\ref{fig_dn}. We observe that at high temperature the
population distribution $P(N_A)$ is well approximated by 
a gaussian distribution, but displays two symmetric
peaks about the central value $N_A=500$ when the system approaches
the critical temperature. Finally, at low temperature we find
two very narrow peaks, well separated and centered around $N_A=950$
and $N_B=50$.

%%%%%%%%%%%%%%%%%%%%%%%%%%%%%%% FIG %%%%%%%%%%%%%%%%
%%%%%%%%%%%%%%%% End FIG %%%%%%%%%%%%%%%%%%%%%

Let us consider more closely the behavior of the order parameter near
the critical point.  According to mean field description of section
II, at $T_{cr}$ the less negative eigenvalue of the matrix of
evolution vanishes. The picture looks somehow different when
microscopic fluctuations are accounted for. One observes a remarkable
behavior in the order parameter evolution, namely in the earlier
regime it grows in a power law fashion as $(N_A-N_B)^2\sim t$, and not
exponentially. To explain this phenomenology we remark that the
evolution of the order parameter near $T_{cr}$ can be assimilated to
that of a particle undergoing a random walk. We assume that the
early evolution of $\Delta N(t)= N_A(t) - N/2$ can be effectively described
by the following equation:

\begin{equation}
\frac{\partial \Delta N(t)}{\partial t} = \gamma(T_b) \Delta N(t) + A(T_b) \eta
\label{ansatz}
\end{equation}

with $\gamma$ lesser than zero above the critical temperature and
greater than zero under the critical temperature and $\eta$ a white
gaussian noise. In this case 

\begin{equation}
 C(\tau,t)=
\langle \Delta N(t+\tau) \Delta N(t) \rangle=
\Big[\langle \Delta N(0) \Delta N(0) \rangle+\frac{A^2}{2 \gamma}\Big]
e^{\gamma(2 t+ \tau)}
-\frac{A^2}{2 \gamma} e^{\gamma \tau}
\end{equation}
which means that if $\gamma<0$ (asymptotically in a time greater than 
$1/|\gamma|$) the equal time correlation function becomes:

\begin{equation}
\langle \Delta N(t) \Delta N(t)  \rangle \to -\frac{A^2}{2\gamma}
\end{equation}

or equivalently $A=\sqrt{-2\gamma \langle \Delta N^2 \rangle }$,
where $\Delta N^2=\lim_{t \to \infty} C(0,t)$.
On the other hand, when $\gamma=0$ (i.e. in correspondence of the critical
point), the equal time correlation function displays a diffusive
behavior:

\begin{equation}
\langle \Delta N(t) \Delta N(t)  \rangle \to A^2 t
\end{equation}
that means that the diffusion coefficient for the 
variable $\Delta N$ is given by $D=A^2/2$.

Eventually in the late regime the growth saturates due to the onsert
of non linear effects.

\begin{figure}[ht]
\centerline {\includegraphics[clip=true,width=8cm,keepaspectratio]{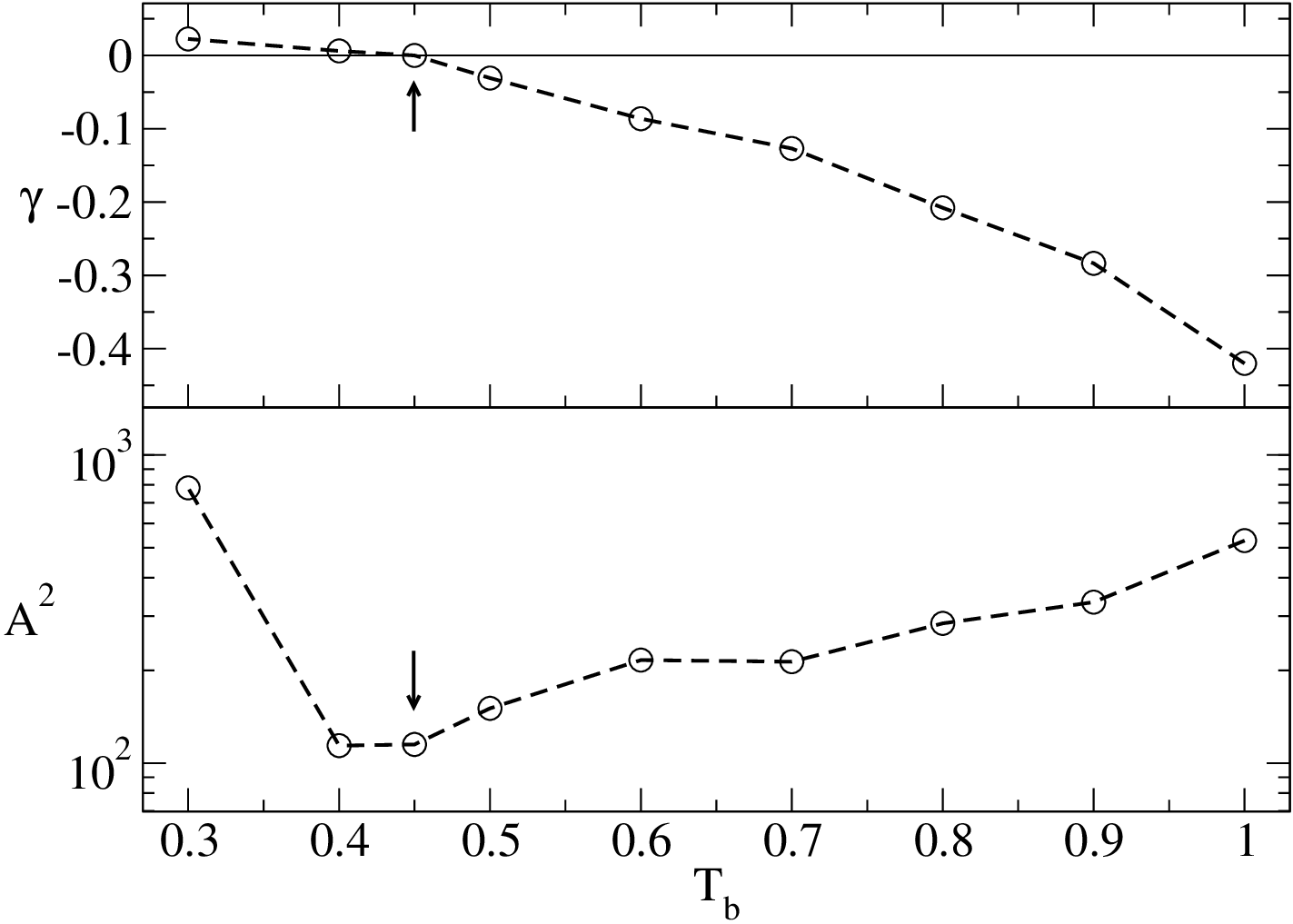}}
\caption{ Study of  $A^2$  and $\gamma$ versus $T_b$.}
\label{fig_crit}
\end{figure}

In order to verify the plausibility of eq.~\eqref{ansatz} 
we determine  $A$ and $\gamma$ by extracting them from numerical 
measures of $\lim_{t \to \infty} C(\tau,t)$ and imposing the
asymptotic behavior $-\frac{A^2}{2 \gamma} e^{\gamma \tau}$.
In fig.~\ref{fig_crit} we display the values of the constants
obtained from simulations above the critical temperature. 
We also added to the plot the value of  the diffusion coefficient, $D=A^2/2$,
as measured at the critical temperature.

%%%%%%%%%%%%%%%%%%%%%%%%%%%%%%%%%%%%%%%%%%%%%%%%%%%%%%%%%5
%%%%%%%%%%%%%%%%%%%%%%%%%%%%%%% FIG crit.eps %%%%%%%%%%%%%%%%
%%%%%%%%%%%%%%%% End crit.eps%%%%%%%%%%%%%%%%%%%%%

%\subsection{Anomalous behavior of the order parameter}

%%%%%%%%%%%%%%%%%%%%%%%%%%%%%%%%%%%%%%%%%%%%%%%%%%%%%%%%%5
%%%%%%%%%%%%%%%%%%%%%%%%%%%%%%% FIG strange.eps %%%%%%%%%%%%%%%%
\begin{figure}[htb]

\centerline {\includegraphics[clip=true,width=8cm,keepaspectratio]
{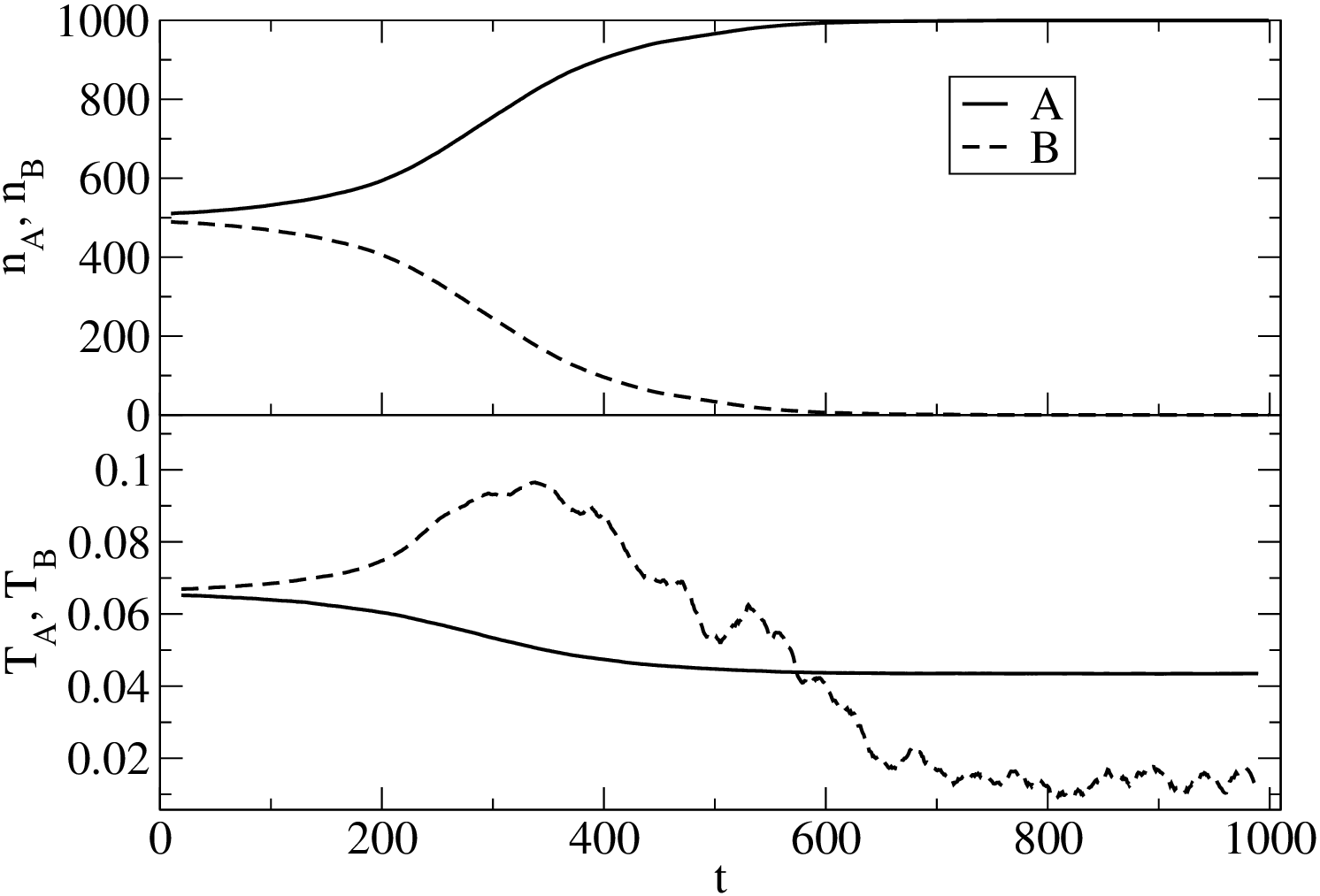}}
\caption
{Populations and granular temperatures
of the two compartments when $\tau_s$ is small. 
Notice that the temperature of the
less populated compartment initially grows, and successively
decreases below the temperature of the other compartment.}
\label{strange}
\end{figure}
%%%%%%%%%%%%%%%% End strange.eps%%%%%%%%%%%%%%%%%%%%%
%%%%%%%%%%%%%%%%%%%%%%%%%%%%%%%%%%%%%%%%%%%%%%%%%%%%%%%%%5

Interestingly, when the characteristic time $\tau_s$ decreases
(namely $\tau_s=0.1$) 
a new phenomenon appears: the temperature of the less populated
compartment exhibits an initial regime, during which it increases,
due to the decrement of the number of collisions the particles experience,
followed by a second regime where the temperature instead {\it decreases}
and eventually reaches a temperature lower than the temperature of the
more populated container.
This kind of anomaly is  due
to the fact that the fastest and more energetic particles are removed 
causing a large negative fluctuation of the associated
average kinetic energy. The positive energy flux provided by the
heat bath is not sufficient to compensate the energy loss
due to the removal of the fast particles.
This phenomenon is not observed in the mean field model, because it
results from the large non-gaussian fluctuations of the velocity PDF.
We checked the fourth cumulant and observed that whereas the cumulant
relative to the populated compartment corresponds to a gaussian,
the cumulant of the small population strongly fluctuates. 
Experimentally it might be hard to observe such a phenomenon, which is
probably an artifact of the model, because the limit of small $\tau_s$
is not very realistic.

Finally, our schematization of the compartmentalized granular
gas recalls the Gibbs ensemble method of equilibrium statistical mechanics,
whereby it is possible to study first order phase coexistence
without interfaces \cite{Frenkel}. 

\section{Conclusions}

To summarize, we introduced a model for a compartmentalized
granular gas which allows to bridge between the
microscopic level to the hydrodynamic level.
In the first part of the
paper we have derived a Fokker-Planck-Boltzmann description starting
from the stochastic evolution of the particles coordinates. Next, by
employing a gaussian ansatz for the velocity distribution function, we
have obtained a closed set of equations for the
slowly varying fields, namely the granular temperatures and occupation
numbers of each compartment.  
Let us comment that with respect to existing
flux models, our approach treats analytically the granular temperature 
on equal footing as the occupation variables. 
The solution of the resulting equations
shows the existence of two different regimes: at strong shakings the
populations in the two compartments are symmetric and for weak
shakings the two populations are asymmetric together with their
granular temperatures. A critical point separates these two
behaviors. The dynamics has been characterized in the various regimes.
In the last part of the paper we have solved the full model by means of
DSMC. The presence of stochastic fluctuations leads to 
properties not observed in a mean field
description. These are the presence of critical fluctuations, 
of anomalies in the dynamics of the populations and in 
population fluctuations.

\section{Acknowledgments}

U.M.B.M. acknowledges the support of the 
Cofin MIUR ``Fisica Statistica di Sistemi Classici e Quantistici''. 
A.P. thanks INFM Center for Statistical Mechanics and Complexity.


\begin{thebibliography}{99}
\bibitem{sonine} In order to improve
the present approximation one could express the 
probability distribution function as a Gaussian multiplied by
a linear combination 
of orthogonal polynomials, called Sonine polynomials.
After expressing the coefficients multiplying the polynomials 
in terms of the moments of the distribution function
one obtains the non gaussian corrections to the
velocity distribution.
The method allows to include the effect of 
fluctuations around the  gaussian. However, for the sake
of simplicity we shall not pursue further this route.
\bibitem{extension}
The self-consistent equations obtained above
can be extended to a network of
compartments in a straightforward fashion. The 
study of such a system will be the subject of future 
work.
\bibitem{facciamoipedanti}
This does not mean
that the transition from the symmetric behavior to the asymmetric behavior
of the solutions is continuous. On the contrary, as we shall show below,
we observe hysteresis, which is consistent with a discontinuous transition.
Such a situation is different from the picture presented in ref.
\cite{Lohse}, based on a flux model, that the transition is continuous
when the compartments are two
and discontinuous when these are more than two.
\end{thebibliography}
\end{document}